\documentclass[10pt]{article}
\usepackage{spconf}

\usepackage{verbatim}
\usepackage{amsfonts,amsmath,mathrsfs,amssymb,amsbsy}
\usepackage[final]{graphicx}
\usepackage{times,cite}
\usepackage{subfig}

\long\def\symbolfootnote[#1]#2{\begingroup%
\def\thefootnote{\fnsymbol{footnote}}\footnote[#1]{#2}\endgroup}

\usepackage{verbatim}
\usepackage[]{float,latexsym}
\usepackage{amsfonts,amsmath,mathrsfs,amssymb,amsbsy}
\usepackage{url}

\newtheorem{theorem}{Theorem}[section]

\newcommand{\Prob}{\mathsf{P}}
\newcommand{\Expect}{\mathsf{E}}

\usepackage{color}
\definecolor{lightblue}{rgb}{.7, .8, 1}
\definecolor{lightgreen}{rgb}{.6, 1, .6}
\usepackage{color}
\definecolor{brown}{rgb}{1,0.38,0.03}

\definecolor{OliveGreen}{rgb}{.2,0.6,0.2}
\definecolor{BrickRed}{rgb}{.7,0.2,0.2}

\newcommand{\ignore}[1]{} %%% {} empty inside
\long\def\symbolfootnote[#1]#2{\begingroup%
\def\thefootnote{\fnsymbol{footnote}}\footnote[#1]{#2}\endgroup}

\DeclareMathOperator*{\argmin}{arg\,min}
\newcommand{\bsp}{\begin{split}}

\newcommand{\esp}{\end{split}}

\begin{document}

\sloppy
\ninept

\title{Wavelet Shrinkage and Thresholding based Robust Classification for Brain-Computer Interface}

\name{Taposh Banerjee$^{\star}$ \; John Choi$^{\dagger}$ \; Bijan Pesaran$^{\dagger}$ \; Demba Ba$^{\star}$ \; and \; Vahid Tarokh$^{\star}$ \thanks{The work at both Harvard and NYU was supported by the Army Research Office MURI Contract Number W911NF-16-1-0368.}}

\address{$^{\star}$ School of Engineering and Applied Sciences, Harvard University\\
    $^{\dagger}$ Center for Neural Science, New York University}
%% Create the title:
\maketitle

\begin{abstract}
A macaque monkey is trained to perform two different kinds of tasks, memory aided and visually aided. In each task, the monkey saccades to eight possible target locations. A classifier is proposed for direction decoding and task decoding based on
local field potentials (LFP) collected from the prefrontal cortex.   	
The LFP time-series data is modeled in a nonparametric regression 
framework, as a function corrupted by Gaussian noise. 
It is shown that if the function belongs to Besov bodies, then using the proposed wavelet shrinkage and thresholding based classifier is robust and consistent. The classifier is then applied to the LFP data to achieve high decoding performance. 
The proposed classifier is also quite general and can be applied for the classification of other types of time-series data as well, 
not necessarily brain data. 
\end{abstract}


\section{Introduction}
In this paper, we consider the problem of robust classification of local field potentials (LFPs).  
A macaque monkey is trained to perform memory aided and visually aided tasks. In each type of task, the monkey is trained to saccade to one of eight possible target locations on a LED illuminated board. LFPs are collected from 
the monkey's brain by inserting a micro-electrode array into its prefrontal cortex. The objective is to decode what type of task (memory aided or visually aided) the monkey is doing, and also where the monkey is looking in each task. 
Successful decoding of brain signals from macaque monkey provides important insights about the human brain, due 
to the similarity of the two brains. These insights will also be fundamental to the design 
of effective brain-computer interfaces. 

Power spectrum based techniques are popular in the neuroscience literature; see \cite{rao2013brain}, \cite{markowitz2011optimizing}, 
and the references therein. 
This is because often the LFP is modeled as a time-series, and in that generality, power spectrum 
based techniques are quite effective. In \cite{BaneFS2017}, we modeled the LFP in a
nonparametric regression framework, as function corrupted by Gaussian noise. We showed that 
if the function is smooth and belongs to a Sobolev class, then using Fourier series based Pinsker or blockwise James-Stein classifiers
lead to consistent and robust classification. These classifiers were applied to LFP data to obtain high decoding accuracy. 

In this paper, we argue that wavelet shrinkage and thresholding based classifiers would also lead to robust and consistent classification, but now over a broader class of functions called Besov bodies. The Sobolev classes are a special case of Besov bodies, and hence due to the adaptive minimaxity property of wavelet thresholding, the wavelet-based classifiers proposed in this paper would work as well as Fourier based methods on Sobolev class of functions used in \cite{BaneFS2017}. This is also verified through application of these classifiers on the LFP data in Section~\ref{sec:Numerical} below. On the 
other hand, if the true function in the regression framework has a sparse representation in say wavelet basis, then the wavelet thresholding 
based classifiers would perform better. These arguments also suggest that if the time-series data can be modeled in a regression framework, then power spectrum based tests are not optimal, and should in general lead to suboptimal performance. 

We apply the wavelet thresholding based classifier to decode the eye movement location with an accuracy of 88\%, same as Pinsker's classifier from \cite{BaneFS2017}. We also decode the type of task with an accuracy of 98\%; see Section~\ref{sec:Numerical}. 
The classifiers we have proposed in this paper are quite general in the sense that it can be applied to any type of time-series data, not just to LFP or brain data. 

\section{Memory and Visually Aided Saccade Experiment}\label{sec:monkeyexpt}
\begin{figure}[ht]
	\center
	\includegraphics[width=8.5cm, height=5cm]{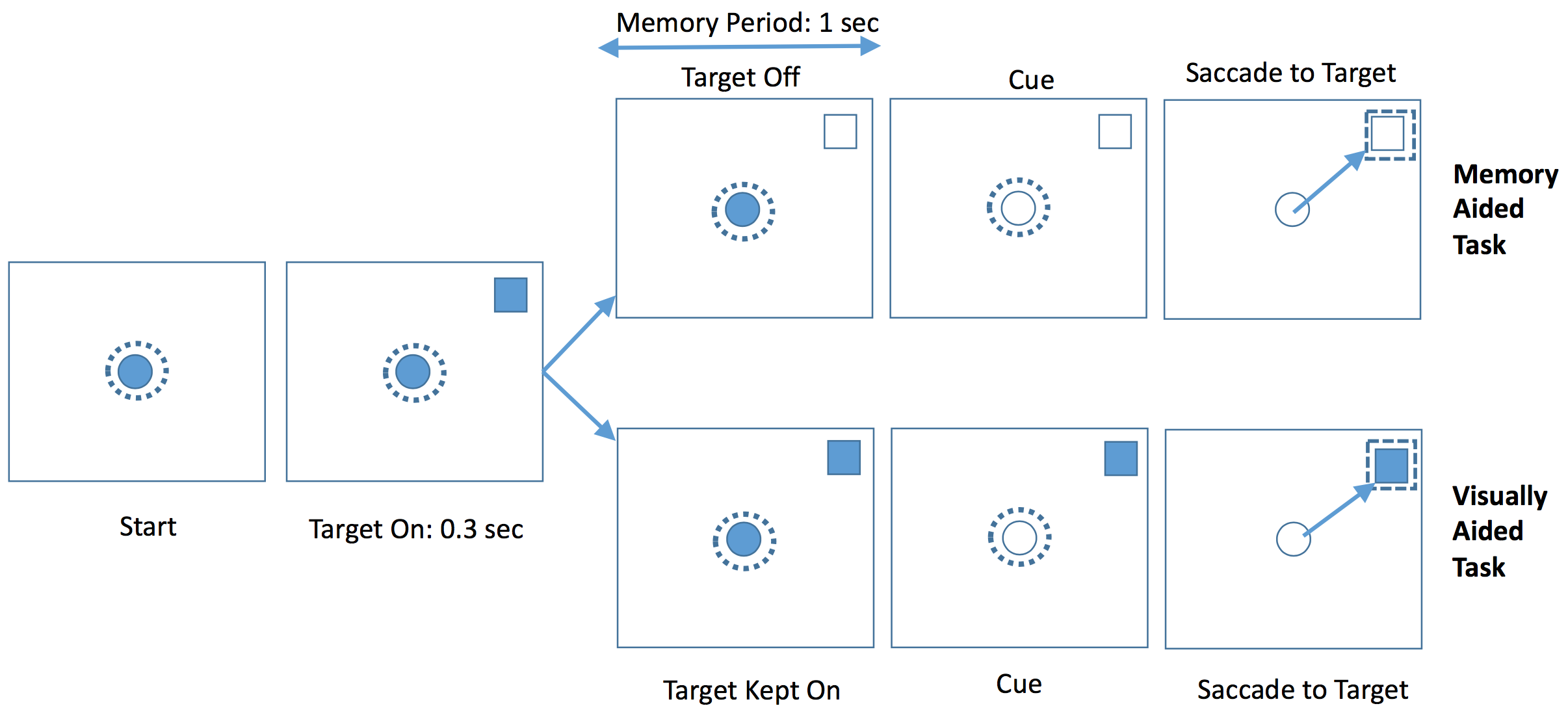}
	\caption{Memory aided and visually aided saccade experiment. The blue or filled square or circle represent an LED illuminated location. Dotted circles and boxes show where the monkey's eyes are fixated. }
		\label{fig:MemVsDelayedSaccade}
\end{figure}

A macaque monkey is trained to perform a series of interleaved memory and visually aided tasks. The visually aided task is also referred to as a delayed task in the following. In both the tasks, 
the task starts with the monkey looking at the illuminated center of a target board. After the monkey's eyes are fixated at the center for a while, a target light
at one of the eight centers (four vertex and four side centers) is switched on for 300 ms. See Fig.~\ref{fig:MemVsDelayedSaccade}. 
The eight target locations are chosen randomly. After this 300 ms length period, the activity in the two tasks starts to differ. In the memory aided task (top flow of boards in Fig.~\ref{fig:MemVsDelayedSaccade}), the target light is switched off for about 1 second. This time corresponds to a memory period in which the monkey has to remember the location of the target. At some later time, a cue is given to the monkey by switching off the center light on the board. After the cue, the monkey is trained to saccade to the target location. The successful saccade to the target marks the end of the memory aided task. In the visually aided task (bottom flow of boards in Fig.~\ref{fig:MemVsDelayedSaccade}), 
the flow is essentially the same, with the important difference that 
the target light is never switched off. The choice between the memory and the delayed task is also random. 
LFP data is collected from electrodes embedded in the cortex of the monkey throughout the experiment. 
The data is collected using a 32 electrode array resulting in 32 parallel streams of LFP data. 
The data used is sampled at $1$ kHz, giving us $1000$ time series samples per channel from the memory period.  In this paper, 
we use the first $500$ samples from the memory period for classification. 
Further details about these tasks can be found in \cite{markowitz2011optimizing}.

There are two statistical inference questions that we want to address:
\begin{enumerate}
\item Predicting where the monkey is looking in a memory-based task. 
\item Classifying what type of task the monkey is doing. 
\end{enumerate}
%In this paper we propose discrete wavelet transform based classification rules, and successfully apply them to the inference problems above. 

\section{A Time Series Classification Problem}\label{sec:Data Modeling}
We model the LFP data, say $\{Y_\ell\}$, using a nonparametric regression framework:
\begin{equation}\label{eq:RegressLFPMod}
Y_\ell = f( \ell / N ) +  Z_\ell, \quad \ell \in \{0, \cdots, N-1\},
\end{equation}
where $f$ belongs to a class $\mathcal{F}$ of square integrable functions and  $Z_\ell \stackrel{\text{i.i.d.}}{\sim}  \mathcal{N}(0,1)$. The idea is that 
$f$ contains the information relevant to the decision making involved, and we model the irrelevant part of the data by noise. 
\begin{figure}[t]
	\center
	\includegraphics[width=9cm, height=6cm]{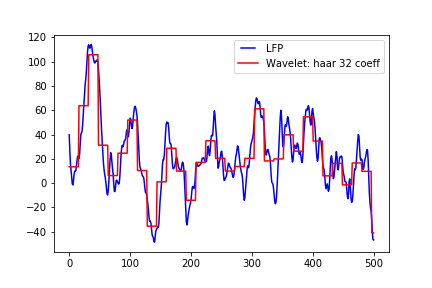}
	\caption{Sample LFP waveform and its approximation using the first 32 Haar wavelet transform coefficients.}
		\label{fig:LFPwithWavelets}
\end{figure}
In Fig.~\ref{fig:LFPwithWavelets}, we show a sample LFP waveform and its reconstruction using first 32 Haar wavelet coefficients. 
The accuracy of the approximation suggests that nonparametric regression provides an appropriate modeling framework for the LFP data. 

We consider the following classification problem:
\begin{equation}\label{eq:classProb}
\begin{split}
H_k: \quad f \in & \; \mathcal{F}_k, \quad k \in \{1, 2, \cdots, m\}, \\
& \mathcal{F}_k \subset \mathcal{F} \subset L_2([0,1]), \quad \mathcal{F}_k \cap \mathcal{F}_j = \emptyset, \mbox{ for } j \neq k.
\end{split}
\end{equation}      
Here, $m=8$ for the target classification problem, and $m=2$ for the type of task classification problem. Further motivation for this classification problem can be found in \cite{BaneFS2017}. 
Our objective is to find 
a classifier that maps the LFP data $\{Y_\ell\}$ to one of $m$ function classes. We need a classifier that works not only on our data set but for a wide variety of similar datasets, collected over time and from different monkeys. Thus, we need a classifier that is robust. 
If $\hat{\delta}(\{Y_\ell\})$ denotes the classifier, then one way to ensure robustness is to seek that the worst case error
\begin{equation}
P_e = \max_{k \in \{1, \cdots, m\}} \;  \sup_{f \in \mathcal{F}_k}  \; \Prob_f(\hat{\delta} \neq k) 
\end{equation}
goes to zero as the number of samples increases. Here, $\Prob_f$ denotes the law of LFP data when the true function is $f$. 
In the next section, we argue that one way to ensure robustness is to choose the minimax estimator over the class $\mathcal{F}$ as a feature for the classifier. 

\section{Using Minimax Estimator as a Feature}\label{sec:minimaxest}
Let $\hat{f}^*_N$ be the minimax estimator for the function class $\mathcal{F}$, i.e., it satisfies as $N\to \infty$
\begin{equation}\label{eq:minimaxest}
\begin{split}
\sup_{f \in \mathcal{F}}\; \Expect_f[ \| \hat{f}^*_N - f \|_2^2] & \to 0\\
\sup_{f \in \mathcal{F}}\; \Expect_f[ \| \hat{f}^*_N - f \|_2^2] &\sim \inf_{\hat{f}} \sup_{f \in \mathcal{F}}\; \Expect_f[ \| \hat{f} - f \|_2^2] ,
\end{split}
\end{equation}
where $\Expect_f$ denotes the expectation with respect to the probability measure of LFP with $f$ as the true function, and 
$\|g\|_2$ is the $L_2[0,1]$ norm. 

Let $\delta_{\text{md}}$ be the minimum distance decoder
\begin{equation}\label{eq:minDistDec}
\delta_{\text{md}}(f) = \argmin_{k \leq  8} \inf_{g \in \mathcal{F}_k}\| f - g\|_2.
\end{equation}
Given the knowledge of the classes $\{\mathcal{F}_k\}$, the decoder $\delta_{\text{md}}$ finds the function $g$ that is closest 
to $f$ across all function classes and picks the index of the class to which $g$ belongs as its value. 

Consider the classifier $\delta_{\text{md}}(\hat{f}^*_N)$, in which the LFP data is first used to obtain a function estimator, 
and then the minimum distance decoder is used to obtain the function class closest to the function estimate. 
We now show that $\delta_{\text{md}}(\hat{f}^*_N)$ is a consistent classifier. 

Let, for $s > 0$, the classes $\{\mathcal{F}_k\}$ be separated by a distance of at least $2 s$:
\begin{equation}\label{eq:2sseperation}
\min_{k \neq  j} \inf_{f \in \mathcal{F}_j, \; g \in \mathcal{F}_k}\| f - g\|_2 \; > \; 2s.
\end{equation}
Then, we have the following theorem. The proof can be found in \cite{BaneFS2017}.

\begin{theorem}
Under the stated assumptions, the maximum decoding error of $\delta_{\text{md}}(\hat{f}^*_N)$ goes to zero:
\begin{equation}\label{eq:consistecyOfMaxErr}
\begin{split}
P_e &= \max_{k \in \{1, \cdots, m\}} \;  \sup_{f \in \mathcal{F}_k}  \; \Prob_f(\delta_{\text{md}}(\hat{f}^*_N) \neq k) \\
&\leq  \frac{1}{s^2}\; \sup_{g \in \mathcal{F}} \;\Expect_g [  \|\hat{f}^*_N - g\|_2^2] \to 0, \quad \mbox{ as } N \to \infty.
\end{split}
\end{equation}
\end{theorem}
Note that the above result is valid for any set of classes, as long as they are well separated. The result does not even depend on 
the number of classes $m$. Thus, it covers both the inference problems we are interested in this paper. 
The maximum decoding or classification error also goes to zero for any decoder $\delta_{\text{md}}(\hat{f}_N)$ for which the maximum estimation error of $\hat{f}_N$ over 
the class $\mathcal{F}$ goes to zero. Among all such estimators, the decay rate for $\delta_{\text{md}}(\hat{f}^*_N)$ is the highest. 

The above result suggests that if we have precise information about the function classes, then we can use the minimax estimator to obtain a consistent and robust classifier. On the other hand, if exact details of the function classes are not known, we can 
use the minimax estimator as a feature to train a robust classifier. A function estimator is an infinite dimensional object, and hence hard 
to use as a feature. In the next section, we discuss estimation in Gaussian sequence models, and how it can help in compactly representing the minimax estimator. 

\section{Estimation in Gaussian Sequence Models}
A Gaussian sequence model is defined as
\begin{equation}\label{eq:GaussSeqMod}
y_I = \theta_I + \epsilon \; z_I, \quad I \in \mathcal{I}.
\end{equation} 
The sequence $\{\theta_I\}$ is the parameter sequence to be estimated using the observations $\{y_I\}$ corrupted by white Gaussian noise sequence $\{z_I\}$, i.e., $z_I \stackrel{\text{i.i.d.}}{\sim} \mathcal{N}(0,1)$, and $\mathcal{I}$ is an index set, and can be one or two dimensional. 

Function estimation in the nonparametric regression model \eqref{eq:RegressLFPMod} is equivalent to parameter estimation in the Gaussian sequence model \eqref{eq:GaussSeqMod}. To understand this equivalence, consider the ideal white noise model (idealized regression model)
\begin{equation}\label{eq:WhiteNoiseMod}
Y(t) = \int_0^t f(s) \; ds + \epsilon \; W_t, \quad t\in [0,1],
\end{equation}
where $W$ is a standard Brownian motion, and $\epsilon > 0$. 
To see the connection of the white noise model \eqref{eq:WhiteNoiseMod} and Gaussian sequence model \eqref{eq:GaussSeqMod}, 
we consider two orthogonal transforms. 
\begin{enumerate}
\item Fourier series: Let $\mathcal{I}=\mathbb{N}$, $\{\phi_k\}$ be the trigonometric series, and 
\begin{equation}\label{eq:GSM_map1}
\begin{split}
y_k &= \int_0^1 \phi_k(t) \; dY(t) \\
&= \int_0^1 \phi_k(t)\; f(t) \; dt + \epsilon \; \int_0^1 \phi_k(t) \; dW(t) \\
&=\theta_k + \epsilon z_k,
\end{split}
\end{equation}
where the integrals with respect to $Y$ and $W$ are stochastic integrals \cite{rogersWilliams}, and 
we have identified
\[
\theta_k = \int_0^1 \phi_k(t) \; f(t) \; dt
\]
and
\[
z_k = \int_0^1 \phi_k(t) \; dW(t)
\]
to get the Gaussian sequence model \eqref{eq:GaussSeqMod}. 
\item Wavelet transform: Let $\phi$ and $\psi$ denote the scaling function (father wavelet) and 
the mother wavelet function, respectively. Also, let $\phi_{jk}$ (similarly $\psi_{jk}$) denote the scaled and shifted version of $\phi$ at level $j$ and shift $k$:  $\phi_{jk}(x) = 2^{j/2} \phi (2^j x -k)$ and $\psi_{jk}(x) = 2^{j/2} \psi (2^j x -k)$. Let us assume that the wavelet basis is also adapted to $L_2[0,1]$, so that for any fixed coarse scale $L$, $\{\phi_{Lk}\}$, $k = 0, 1, \cdots, 2^L-1$, and $\{\psi_{jk}\}$, 
$j \geq L, k=0, 1, \cdots, 2^j-1$, form an orthonormal wavelet basis for $L_2[0,1]$. Now define $\mathcal{I} \subset \mathbb{N}^2$, 
and similar to the Fourier series case, treat $y_I$ as the projection of function $Y(t)$ on the orthonormal basis functions, and $\theta_I$ as the projection of function $f(t)$ on the orthonormal basis functions to recover the Gaussian sequence model \eqref{eq:GaussSeqMod}. 
\end{enumerate}

If we are interested in the mean square error criterion, then due to Hilbert space isomorphism we have 
\begin{equation}\label{eq:MSEGaussMod}
\begin{split}
\text{MSE}(\hat{f}) &= \Expect_f \left[\int_0^1 (f(t) - \hat{f}(\{Y(\cdot)\}; t))^2 \; dt \right] \\
&=  \Expect_\theta \left[ \| \theta - \hat{\theta}(\{y_\mathcal{I}\}) \|_2^2 \right] = \text{MSE}(\hat{\theta}), \mbox{ as } N \to \infty,
\end{split}
\end{equation}
where $\hat{\theta}(\{y_I\})$ are the wavelet coefficients of $\hat{f}$, and can be treated as an estimator of $\{\theta_I\}$ based on the observations $\{y_I\}$. 
Thus, the estimation of function $f$ in the white noise model \eqref{eq:WhiteNoiseMod} is equivalent to the estimation of $\theta_I$ in the \eqref{eq:GaussSeqMod}. Specifically, corresponding to a function class $\mathcal{F}$, define a parameter set  
\begin{equation}\label{eq:ThetaF}
\Theta(\mathcal{F}) = \{\theta: \theta \mbox{ is Fourier or wavelet transform coeff. of} f \in \mathcal{F}\}.
\end{equation}
Then we have
\begin{equation}\label{eq:GSMWhiteNoiseEq}
\begin{split}
\inf_{\hat{f}} \sup_{f \in \mathcal{F}} & \; \Expect_f \left[\int_0^1 (f(t) - \hat{f}(\{Y(\cdot)\}; t))^2 \; dt \right] \\
&= \inf_{\hat{\theta}} \sup_{\theta \in \Theta(\mathcal{F})} \Expect_\theta \left[ \| \theta - \hat{\theta}(\{y_\mathcal{I}\}) \|_2^2 \right].
\end{split}
\end{equation}
Thus, to obtain the optimal function estimator, we take the Fourier series or wavelet transform of the function $Y(t)$ in \eqref{eq:WhiteNoiseMod}, and apply the optimal parameter estimator to the transform coefficients. Finally, 
we reconstruct a function from the optimal estimator. 

It can be shown that the regression model \eqref{eq:RegressLFPMod} is also equivalent to both the white noise model and the Gaussian sequence model. The equivalence can be obtained between the regression model and the Gaussian sequence model, 
by treating $y_k$ as approximate Fourier series coefficients or by treating $y_I$ as the discrete wavelet transform coefficients, both obtained from the time series data $\{Y_k\}$, and then taking $N \to \infty$. For a more detailed discussion on these mappings, see \cite{Johnstone2015Book} or \cite{tsybakov2009introduction}. 

\subsection{Minimax Estimation on Besov Bodies}
We now state a major result on minimax estimation in Gaussian sequence model that has implications for estimation in white noise model. Because of the asymptotic equivalence between these models and the regression model, the estimator obtained is approximately minimax for the regression model as well. 

Consider the white noise model \eqref{eq:WhiteNoiseMod}, and let $\theta_{jk}(f)$ denote the wavelet coefficients for a function $f \in L_2[0,1]$ at level $j$ and at time shift $k$. Also let $\theta(f) = \{\theta_{jk}(f)\}$.
We use $\theta_{j.}$ to denote wavelet coefficients at level $j$. Let $\alpha >0$, $p \in (0, \infty]$, $q \in (0,\infty]$, and define the Besov bodies with $a = \alpha -1/2 -1/p$, 
\begin{equation}\label{eq:BesovBodies}
\Theta_{p,q}^\alpha(C) = \{\theta: \sum_j 2^{ajq} \; \| \theta_{j.} \|_p^q  \; \leq \; C^q\}.
\end{equation}
Further, define the function spaces
\begin{equation}\label{eq:FuncSpBesov}
\mathcal{F}_{p,q}^\alpha(C) = \{f: f\in L_2[0,1] \mbox{ and } \theta(f)\in \Theta_{p,q}^\alpha(C)  \}.
\end{equation}
Apply wavelet transform to the white noise model to obtain the Gaussian sequence model
\begin{equation}
y_{jk} = \theta_{jk} + \epsilon \; z_{jk}.
\end{equation} 
Clearly, we have the equivalence \eqref{eq:GSMWhiteNoiseEq}. 

We now define a soft-thresholding estimator. Let $J = \log \epsilon^{-2}$ and $\lambda_\epsilon = \sqrt{2 \log \epsilon^{-2}}$, and 
define 
\begin{equation}\label{eq:SoftThre}
\hat{\theta}^S_{jk}  = 
   \begin{cases}
      \delta_S(y_{jk}, \lambda_\epsilon \epsilon), & \text{if}\ j \leq J  \\
      0, & \text{otherwise},
    \end{cases}
\end{equation}
where $\delta_S(x, \lambda) = sign(x) \; (|x| - \lambda)_+$.
We have the following theorem. 
\begin{theorem}[\cite{Johnstone2015Book}]
Let $r = 2\alpha/(2\alpha+1)$, then for any Besov body, as $\epsilon \to 0$,
\begin{equation}\label{eq:thmbesov}
\begin{split}
\sup_{\theta \in \Theta_{p,q}^\alpha(C)} \Expect_\theta \left[ \| \theta - \hat{\theta}^S \|_2^2 \right] &\sim 
\inf_{\hat{\theta}} \sup_{\theta \in \Theta_{p,q}^\alpha} \Expect_\theta \left[ \| \theta - \hat{\theta} \|_2^2 \right] \\
&\sim C^{2 (1-r)} \epsilon^{2r}.
\end{split}
\end{equation}
\end{theorem}
In words, the theorem implies that if the observations are from the white noise model \eqref{eq:WhiteNoiseMod}, then
the soft thresholding wavelet estimator \eqref{eq:SoftThre} are adaptively asymptotically minimax for any Besov body, i.e., for any choice of $p, q, \alpha$. As mentioned earlier, this implies also that soft thresholding estimator applied to the discrete wavelet transform
coefficients of the time-series data, are also asymptotically minimax for the regression model \eqref{eq:RegressLFPMod}. 
Note that the Besov bodies cover a wide range of smoothness, including the Sobolev spaces for which Pinsker's estimator 
used in \cite{BaneFS2017} is optimal. Thus, the estimator based on wavelets is expected to perform as well as those based on Fourier
when the true function is in the Sobolev class. Moreover, for spaces with sparse representation, $p < 2$, the wavelet thresholding perform better.

\section{An LFP Classification Algorithm}
We now propose an algorithm for the LFP classification problems discussed in Section~\ref{sec:Data Modeling}.
Recall that we argued in Section~\ref{sec:minimaxest} that using minimax estimator as features would lead to robust and consistent 
classification. In the previous section, we discussed the structure of minimax estimator over Besov bodies. This leads us to the following classifier. 

\begin{enumerate}
\item \textit{Fix parameters}: Fix $J$, $\lambda$, a parameter $T$, and choice of wavelet Haar, Daubechies, etc.
\item \textit{Compute DWT coefficients}: Use the LFP time-series to compute $N$ discrete wavelet transform coefficients $\{y_{jk}\}$. 
\item \textit{Use Shrinkage and Thresholding}:  Compute a version of minimax estimator
\begin{equation}
\hat{\theta}^S_{jk}  = 
   \begin{cases}
   y_{jk}& \text{if}\ j \leq T  \\
      \delta_S(y_{jk}, \lambda), & \text{if}\ T < j \leq J  \\
      0, & \text{otherwise},
    \end{cases}
\end{equation}
and use $\hat{\theta}^S$ as features. Keep the coarsest coefficients intact. 
\item \textit{Dimensionality Reduction (optional)}: If LFP data is collected using multiple channels, then use principal component analysis to project the $\hat{\theta}^S$ coefficients from all the channels (vectorized into a single high-dimensional vector) into a $P$ dimensional subspace. 
\item \textit{Train LDA}: Train a linear discriminant analysis classifier.
\item \textit{Cross-validation}: Estimate the generalization error using cross-validation. 
\item \textit{Optimize free parameters}: Optimize over the choice of $J$, $T$, $\lambda$, $P$ and the choice of wavelet. 
\end{enumerate}

\section{Numerical Results}\label{sec:Numerical}
The classification algorithm from the previous section was applied to the LFP data collected. 
The results are based on data from $1467$ trials carried out across $9$ recording sessions, out of which $736$ were from memory based tasks and the rest were from visually aided tasks. LFPs were collected from $32$ channels. The depths 
of the electrodes were the same across all the trials. From each trial, LFP data from the memory period (see Fig.~\ref{fig:MemVsDelayedSaccade}) of length $500$ (giving a total of $500 \times 32$ data points per trial) was used to obtain a sample.  The performance was evaluated using leave-one-session-out cross-validation. 

\begin{figure}[ht]
	\center
	\includegraphics[width=7cm, height=5cm]{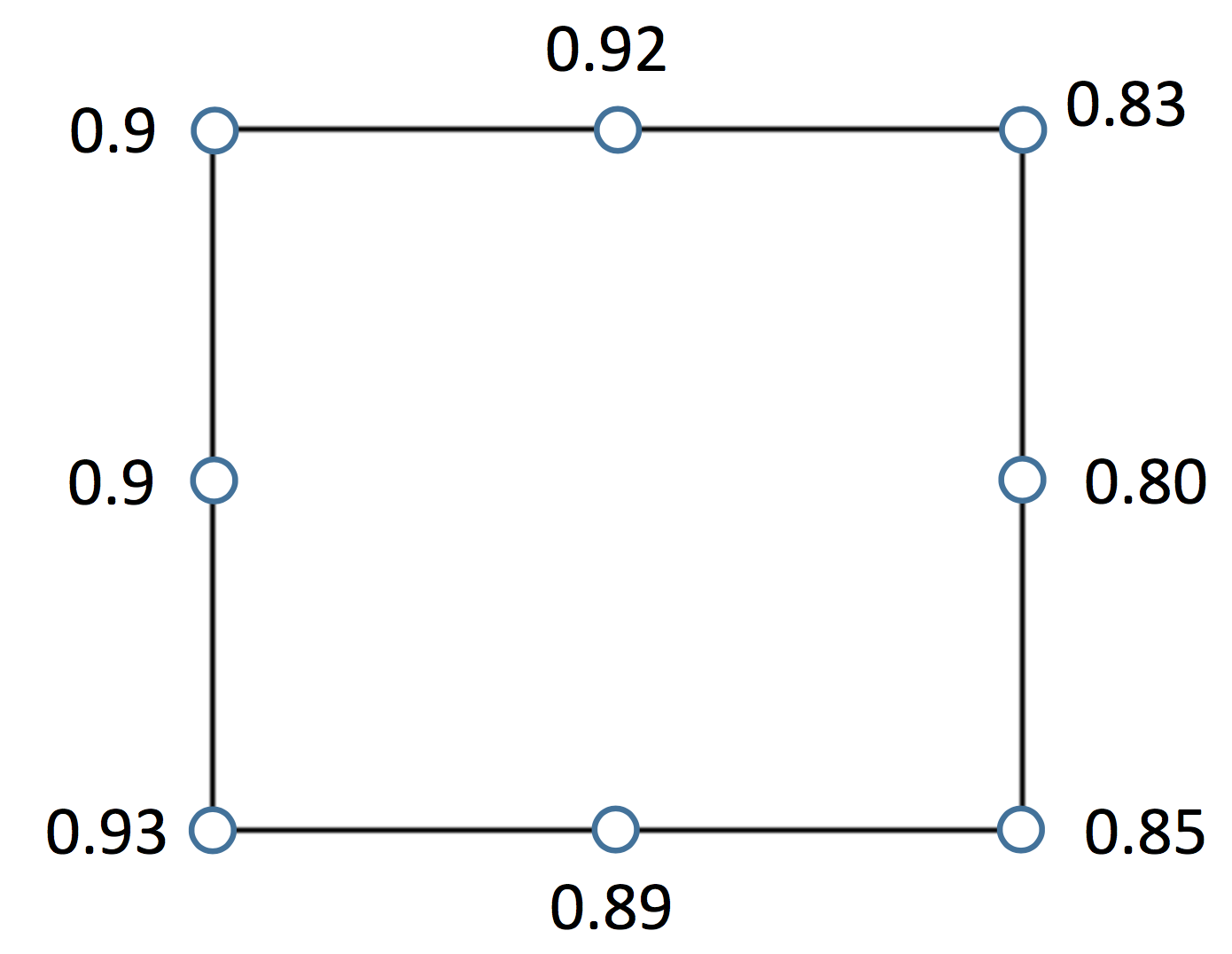}
		\caption{Classification accuracy (conditional probability of correct decoding) per target for the eight targets. The average accuracy is $88$\% for the Wavelet-based classifier, same as Pinsker's classifier from \cite{BaneFS2017}.}
		\label{fig:DecodingPerformance}
\end{figure}

In Fig.~\ref{fig:DecodingPerformance}, 
we show the performance of the wavelet-based classifier for direction decoding in $736$ memory-based tasks. 
The average decoding performance is $88\%$, same as Fourier series based Pinsker's classifier from \cite{BaneFS2017}. The performance reported is for Haar wavelets, and using the parameters $T=J=20$, and $P=200$.
We note that it is shown in \cite{BaneFS2017} that Pinsker's classifier performs better than power spectrum based techniques, 
where absolute values of coefficients are used. 

In Table~\ref{tab:MemVsDel}, we show the performance for decoding the type of task (memory vs delayed). The decoding accuracy on an average is $98.5 \%$. The parameter $T$, $J$, and $P$ used were the same as above. 
\begin{table}[htbp]
  \centering
  \caption{Performance of wavelet-based classifier for decoding type of task. Average decoding performance is $0.985$.}
  \label{tab:MemVsDel}
  \begin{tabular}{| c | c |}
    \hline
    Memory & Delay\\
    \hline
    0.99 & 0.98\\
    \hline
  \end{tabular}
\end{table}

\section{Conclusions and Future Work}
We proposed a wavelet thresholding based robust and consistent classifier for time-series data and applied it to brain data to achieve high decoding accuracy. In future, we will apply the classifier to LFP data collected from other monkeys to test classifier's robustness. 
We will also consider the classification of other time-series data, not just brain data.

%\vspace{-0.2cm}
%\footnotesize
%\pagebreak
%\pagebreak
%\newpage
%\newpage
%\nocite{*}
\bibliographystyle{ieeetr}

%\bibliographystyle{elsarticle-harv}

%\bibliographystyle{elsarticle-harv}
%\bibliography{QCD_verVV}
%\newpage

\bibliography{QCD_verSubmitted}

\end{document}